\documentclass[cpp,fleqn]{w-art}
\usepackage{times}
\usepackage{w-thm}
\usepackage[]{graphicx}
\begin{document}
\DOIsuffix{theDOIsuffix}
\Volume{42}
\Issue{1}
\Month{01}
\Year{2003}
\pagespan{3}{}



\title[Hierarchy of FDTs for the OCP]{Hierarchy of Fluctuation-Dissipation Theorems for the Classical One-Component Plasma}


\author[K.I. Golden]{Kenneth I. Golden\footnote{E-mail: {\sf kgolden@uvm.edu}}} \address[]{Department of Mathematics and Statistics \& Department of Physics, University of Vermont, Burlington, VT 05405}
\author[J.T. Heath]{Joshuah T. Heath\footnote{Email: {\sf jtheath@uvm.edu}}}
\address[]{}
\begin{abstract}
We have derived tractable cubic $(p=3)$ and quartic $(p=4)$ fluctuation-dissipation theorems (FDTs) for the classical one-component plasma in a form that links a single $(p+1)$-point dynamical structure function to a linear combination of $p$th-order density response functions amenable to calculation from model plasma kinetic equations. For $p\ge 3$, we note the emergence of "remainder" contributions comprised of clusters of lower-order dynamical structure functions which can be ultimately traded for response functions vis-\'a-vis the linear and quadratic FDTs. Our analysis provides insight into the structure of the FDT hierarchy.
\end{abstract}
\maketitle                   





\noindent \section{Introduction}

\indent Over the past six decades, the Fluctuation-Dissipation Theorem (FDT) has become a powerful tool in statistical physics, thanks in large part to the pioneering derivation of R. Kubo [1].  The well-known conventional FDT provides a link between the linear response of a system to a weak external perturbation and equilibrium two-point correlations of the system’s fluctuating quantities. However, the system response need not be restricted to be linear.  In the family of nonlinear response functions, the properties and explicit functional forms of quadratic and cubic response functions have been extensively studied in condensed matter physics [2], plasma physics [3], and nonlinear optics [4-6].  

The natural extension of the Kubo formalism leads to the notion of the hierarchy of FDTs wherein each member FDT establishes the relation between its $p$th-order response functions and their companion $(p+1)$-point $(p\ge 1)$ equilibrium correlations of fluctuating quantities.  This is a topic that has been studied by scientists representing a wide range of disciplines, most notably, plasma physics [7-9], nonlinear optics [5, 10], chemistry [11, 12], and statistical physics [13-16].

The conventional nonlinear FDTs, in their most commonly accepted yet most primitive forms, link the pth-order response function to a combination of $(p+1)$-point correlation functions interfering with each other through their entangled Liouville space paths.  For the quadratic FDT featuring three-point correlation functions, this means that two of the three local number or current density operators are nested inside of Poisson brackets (classical FDTs) or commutators (quantum FDTs), and, as such, are not so easily amenable to computation.  This formalism becomes all the more unwieldy for higher-order FDTs that feature Poisson brackets (commutators) nested inside Poisson brackets (commutators), etc.

The main goal of this paper is to develop a procedure that entirely circumvents the issue of the nested Poisson brackets encountered in the Kubo formalism leading ultimately to compact and tractable cubic $(p=3)$ and quartic $(p=4)$  FDTs, in which a single $(p+1)$- point dynamical structure function, now free of interference in the Liouville space paths, is expressed as a linear combination of the $p$th-order density response functions, which, in turn, can be readily calculated from model equations of motion. The basic ingredients in the procedure are i) the hierarchy of static FDTs, developed in Sec. 3 through $p=4$; ii) the structure of the rigorously established dynamical quadratic FDT [7(a),(b)]; iii) the invariance of the  nonlinear dynamical FDT under permutations of its wave vector-frequency arguments; and iiii) the generalized Poincare-Bertrand theorems. Our study is carried out for the classical OCP subjected to a longitudinal scalar potential perturbation.  

The plan of the paper is as follows: The relevant OCP response and structure functions are introduced in Sec. 2.  In Secs. 3 and 4 we formulate the cubic and quartic static and dynamical FDTs.  Concluding remarks follow in Sec. 5.  
\\
\section{Static Response and Structure Functions}

Consider a collection of N classical point ions, each of mass m and carrying charge Ze, immersed in a uniform neutralizing background of degenerate rigid electrons; the entire system occupies the large but bounded volume $V$, with $n_0=N/V$ being the average density. The microscopic particle density and its spatial Fourier transform are given by
\\
\begin{align}
n({\bf r},\,t)=\sum_i \delta({\bf r}-{\bf x}_i(t)),\qquad n({\bf k},\,t)=\sum_i \exp(-i {\bf k}\cdot {\bf x}_i (t))\label{1}
\end{align}
\\
Addressing first the derivation of the static FDT hierarchy, we suppose that the equilibrium system consists of the N OCP particles subjected to an external Coulomb potential $\hat{\Phi}({\bf r})=\hat{Q}/r$ (originating from a weak external charge $\hat{Q}$ located at the origin); the potential of the external force acting on ion i is $\hat{U}({\bf x}_i)=Ze \hat{\Phi}({\bf x}_i)$. A single such potential produces density excitations (to all orders in $\hat{U}$). The latter are linked to the former by wave vector dependent density response functions defined through the hierarchy of constitutive relations
\\
\begin{align}
\langle n({\bf k})\rangle^{(p)}=\frac{1}{V^{p-1}}\sum_{{\bf k}_1}\sum_{{\bf k}_2}...\sum_{{\bf k}_p} \hat{\chi}({\bf k}_1,\,{\bf k}_2,\,...,\,{\bf k}_p)\hat{U}({\bf k}_1)\hat{U}({\bf k}_2)...\hat{U}({\bf k}_p)\delta_{{\bf k}_1+{\bf k}_2+...+{\bf k}_p-{\bf k}}\label{2}
\end{align}
\\
The angular brackets denote an ensemble-averaged quantity: $\langle ...\rangle^{(p)}=O(\hat{U}^p)$ refers to ensemble averaging over the perturbed system for $p\ge 1$ and over the unperturbed system for $p=0$.

We wish to establish the relation between $\hat{\chi}({\bf k}_1,\,{\bf k}_2,\,...{\bf k}_p)$ and the $(p+1)$- point static structure function
\\
\begin{align}
&NS_{p+1}({\bf k}_1,\,{\bf k}_2,\,...{\bf k}_p)\delta_{{\bf k}_1+{\bf k}_2+...+{\bf k}_p-{\bf k}}=\langle n({\bf k}_1) n({\bf k}_2)...n({\bf k}_p)n(-{\bf k})\rangle^{(0)};\label{3} \\
&\qquad \qquad\qquad \qquad \qquad \qquad \qquad \qquad  \,\,\,({\bf k}_1\not =0,\,{\bf k}_2\not =0,\,...,\,{\bf k}_p\not =0)\notag
\end{align}
\\
The definition (3) is in keeping with the customary
$( p + 1)$ -point structure function notation featuring only the p linearly independent
wave vectors; indeed, any set of $p$ linearly independent wave vectors would suffice for the description of the structure function. This latter is borne out, for example, by the permutation symmetries governing $S_4$ and $S_5$:
 \\
\begin{align}
S_4({\bf k}_1,\,{\bf k}_2,\,{\bf k}_3)=S_4(-{\bf k},\,{\bf k}_1,\,{\bf k}_2)=S_4({\bf k}_3,\,-{\bf k},\,{\bf k}_1)=S_4({\bf k}_2,\,{\bf k}_3,\,-{\bf k});\quad {\bf k}={\bf k}_1+{\bf k}_2+{\bf k}_3\label{4}
\end{align}
\begin{align}
S_5({\bf k}_1,\,{\bf k}_2,\,{\bf k}_3,\,{\bf k}_4)=S_5(-{\bf k},\,{\bf k}_1,\,{\bf k}_2,\,{\bf k}_3)&=S_5({\bf k}_4,\,-{\bf k},\,{\bf k}_1,\,{\bf k}_2)=S_5({\bf k}_3,\,{\bf k}_4,\,-{\bf k},\,{\bf k}_1)\notag\\
&=S_5({\bf k}_2,\,{\bf k}_3,\,{\bf k}_4,\,-{\bf k});\,\,\,\,\,\,\,\,\,\,\,\,{\bf k}={\bf k}_1+{\bf k}_2+{\bf k}_3+{\bf k}_4\label{5}
\end{align}
\\
\section{Hierarchy of Static Fluctuation-Dissipation Theorems}
The starting point for the derivation of the static FDTs is the equilibrium Liouville density:
\\
\begin{subequations}
\begin{equation}
\Omega=\frac{\exp(-\beta H)}{\int d\Gamma \exp(-\beta H)}\label{6a}
\end{equation}
\begin{equation}
\Omega^{(0)}=\frac{\exp(-\beta H^{(0)})}{\int d\Gamma \exp(-\beta H^{(0)})}\label{6b}
\end{equation}
\end{subequations}
where
\begin{align}
H=H^{(0)}+\hat{H};\label{7}
\end{align}
\begin{subequations}
\begin{equation}
H^{0}=\underbrace{\sum_{i=1}^N \frac{p_i^2}{2m}}_{\begin{subarray}{1}\textrm{kinetic energy}\\ \textrm{ of particles}\end{subarray}}
+\underbrace{\frac{1}{2}e^2 \sum_{i,j=1,\,i\not=j}^N\phi_{ij}({\bf x}_i-{\bf x}_j)}_{\begin{subarray}{1}\textrm{particle-particle}\\ \textrm{ interactions}\end{subarray}}+H_{pb}+H_{bb}\label{8a}
\end{equation}
\begin{equation}
\hat{H}=\sum_{i=1}^N \hat{U}({\bf x}_i)=\frac{1}{V}\sum_{{\bf k}'\not =0}\hat{U}({\bf k}')n(-{\bf k}')\label{8b}
\end{equation}
\end{subequations}
where $H_{pb}$ describes the plasma-background interaction and $H_{bb}$ describes the background-background interaction, the $(0)$ superscript refers to the unperturbed OCP, and $d\Gamma$ is a differential volume element in the 6N-dimensional phase space. Note the deletion of the divergent $\hat{U}({\bf k}')=0$ component from \eqref{8b}.

The routine calculation of the average density response to arbitrary order in $\hat{U}$ carried out first by expanding the numerator and denominator exponentials of \eqref{6a} in powers of $\hat{H}$:
\\
\begin{align}
\Omega=\Omega^{(0)} \frac{1-\beta \hat{H}+\frac{1}{2} \beta^2 \hat{H}^2-\frac{1}{6}\beta^3\hat{H}^3+...}{1-\beta \langle \hat{H}\rangle^{(0)}+\frac{1}{2}\beta^2 \langle \hat{H}^2\rangle^{(0)}-\frac{1}{6}\beta^3 \langle \hat{H}^3\rangle^{(0)}+...}\label{9}
\end{align} 
with
\begin{align}
\langle \hat{H}\rangle^{(0)}=\frac{1}{V}\sum_{{\bf k}'\not =0} \hat{U}({\bf k}')\langle n(-{\bf k}')\rangle^{(0)}=n_0 \sum_{{\bf k}'\not =0}\hat{U}({\bf k}')\delta_{{\bf k}'}=0\label{10}
\end{align}
The Liouville densities
\begin{align}
&\Omega^{(1)}=-\beta \Omega^{(0)}\hat{H}\label{11}\\
&\Omega^{(2)}=\frac{1}{2}\beta^2 \Omega^{(0)}\left[\hat{H}^2-\langle \hat{H}^2\rangle^{(0)}\right]\label{12}\\
&\Omega^{(3)}=-\frac{1}{6}\beta^3\Omega^{(0)}\left[\hat{H}^3-3\hat{H}\langle \hat{H}^2\rangle^{(0)}-\langle \hat{H}^3\rangle^{(0)}\right]\label{13}\\
&\Omega^{(4)}=\frac{1}{24}\beta^4 \Omega^{(0)}\left[ \hat{H}^4+6\langle \hat{H}^2\rangle^{(0)}\langle \hat{H}^2\rangle^{(0)}-6\hat{H}^2\langle \hat{H}^2\rangle^{(0)}-4\hat{H}\langle \hat{H}^3\rangle^{(0)}-\langle \hat{H}^4\rangle^{(0)}\right]\label{13}
\end{align}
follow from the further development of \eqref{9} in powers of $\hat{H}$, whence from \eqref{8b},
\begin{align}
&\Omega^{(1)}=-\frac{\beta \Omega^{(0)}}{V}\sum_{{\bf k}_1\not =0}\hat{U}({\bf k}_1)n(-{\bf k}_1)\label{15} 
\end{align}
\begin{align}
&\Omega^{(2)}=\frac{\beta^2}{2V^2}\Omega^{(0)}\sum_{{\bf k}_1\not =0,\,{\bf k}_2\not =0}\hat{U}({\bf k}_1)\hat{U}({\bf k}_2)\left[n(-{\bf k}_1)n(-{\bf k}_2)-\langle n(-{\bf k}_1)n(-{\bf k}_2)\rangle^{(0)}    \right]\label{16}
\end{align}
\begin{align}
&\Omega^{(3)}=-\frac{\beta^3}{6V^3}\Omega^{(0)}\sum_{{\bf k}_1,\,{\bf k}_2,\,{\bf k}_3\not =0} \hat{U}({\bf k}_1)\hat{U}({\bf k}_2)\hat{U}({\bf k}_3)\bigg[ n(-{\bf k}_1)n(-{\bf k}_2)n(-{\bf k}_3)-n(-{\bf k}_1)\langle n(-{\bf k}_2)n(-{\bf k}_3)\rangle^{(0)}\notag\\
&-n(-{\bf k}_3)\langle n(-{\bf k}_1)n(-{\bf k}_2)\rangle^{(0)}-n(-{\bf k}_2)\langle n(-{\bf k}_3)n(-{\bf k}_1)\rangle^{(0)}-\langle n(-{\bf k}_1)n(-{\bf k}_2)n(-{\bf k}_3)\rangle^{(0)}\bigg]\label{17}
\end{align}
\begin{align}
&\Omega^{(4)}=\frac{\beta^4}{24V^4} \Omega^{(0)}\sum_{{\bf k}_1,\,{\bf k}_2,\,{\bf k}_3,\,{\bf k}_4\not =0} \hat{U}({\bf k}_1)\hat{U}({\bf k}_2)\hat{U}({\bf k}_3)\hat{U}({\bf k}_4)\bigg[n(-{\bf k}_1)n(-{\bf k}_2)n(-{\bf k}_3)n(-{\bf k}_4)\notag\\
&+2\langle n(-{\bf k}_1)n(-{\bf k}_2)\rangle^{(0)}\langle n(-{\bf k}_3)n(-{\bf k}_4)\rangle^{(0)}
+2\langle n(-{\bf k}_1)n(-{\bf k}_3)\rangle^{(0)}\langle n(-{\bf k}_2)n(-{\bf k}_4)\rangle^{(0)}\notag\\
&+2\langle n(-{\bf k}_1)n(-{\bf k}_4)\rangle^{(0)}\langle n(-{\bf k}_2)n(-{\bf k}_3)\rangle^{(0)}
-n(-{\bf k}_1)n(-{\bf k}_2)\langle n(-{\bf k}_3)n(-{\bf k}_4)\rangle^{(0)}\notag\\
&-n(-{\bf k}_2)n(-{\bf k}_3)\langle n(-{\bf k}_1)n(-{\bf k}_4)\rangle^{(0)}-n(-{\bf k}_1)n(-{\bf k}_3)\langle n(-{\bf k}_2)n(-{\bf k}_4)\rangle^{(0)}\notag\\
&-n(-{\bf k}_2)n(-{\bf k}_4)\langle n(-{\bf k}_1)n(-{\bf k}_3)\rangle^{(0)}
-n(-{\bf k}_1)n(-{\bf k}_4)\langle n(-{\bf k}_2)n(-{\bf k}_3)\rangle^{(0)}\notag\\
&-n(-{\bf k}_3) n(-{\bf k}_4)\langle n(-{\bf k}_1)n(-{\bf k}_2)\rangle^{(0)}
-n(-{\bf k}_1)\langle n(-{\bf k}_2) n(-{\bf k}_3)n(-{\bf k}_4)\rangle^{(0)}\notag\\
&-n(-{\bf k}_2)\langle n(-{\bf k}_1)n(-{\bf k}_3)n(-{\bf k}_4)\rangle^{(0)}
-n(-{\bf k}_3)\langle n(-{\bf k}_2)n(-{\bf k}_1)n(-{\bf k}_4)\rangle^{(0)}\notag\\
&-n(-{\bf k}_4)\langle n(-{\bf k}_1)n(-{\bf k}_2)n(-{\bf k}_3)\rangle^{(0)}
-\langle n(-{\bf k}_1) n(-{\bf k}_2)n(-{\bf k}_3)n(-{\bf k}_4)\rangle^{(0)}\label{18}
\bigg]
\end{align}
The calculation of the average density response 
\begin{align}
\langle n({\bf k})\rangle^{(p)}=\int d\Gamma \Omega^{(p)} n({\bf k}),\quad (p=1,\,2,\,3,\,4)\label{19}
\end{align}
follows by substituting \eqref{15}-\eqref{18} into \eqref{19} and then trading the emergent $\langle nn...n\rangle^{(0)}$ equilibrium density correlation functions for their companion static structure functions via \eqref{3}. Subsequent comparison with the constitutive relations \eqref{2} results in the first four equations of the static FDT hierarchy:
\begin{align}
&\hat{\chi}({\bf k}_1)=-\beta n_0 S_2({\bf k}_1)\label{20}\\
&\hat{\chi}({\bf k}_1,\,{\bf k}_2)=\frac{\beta^2 n_0}{2!}S_3({\bf k}_1,\,{\bf k}_2)\label{21}\\
&\hat{\chi}({\bf k}_1,\,{\bf k}_2,\,{\bf k}_3)=-\frac{\beta^3 n_0}{3!}\bigg( S_4({\bf k}_1,\,{\bf k}_2,\,{\bf k}_3)-R_4({\bf k}_1,\,{\bf k}_2,\,{\bf k}_4)\bigg)\label{22}\\
&\hat{\chi}({\bf k}_1,\,{\bf k}_2,\,{\bf k}_3,\,{\bf k}_4)=\frac{\beta^4 n_0}{4!}\bigg( S_5({\bf k}_1,\,{\bf k}_2,\,{\bf k}_3,\,{\bf k}_4)-R_5({\bf k}_1,\,{\bf k}_2,\,{\bf k}_3,\,{\bf k}_4)\bigg)\label{23}
\end{align}
where
\begin{align}
R_4({\bf k}_1,\,{\bf k}_2,\,{\bf k}_3,\,{\bf k}_4)=NS_2({\bf k}_1)S_2({\bf k}_2)\delta_{{\bf k}-{\bf k}_1}+NS_2({\bf k}_3)S_2({\bf k}_1)\delta_{{\bf k}-{\bf k}_3}+&NS_2({\bf k}_2)S_2({\bf k}_3)\delta_{{\bf k}-{\bf k}_2};\notag\\
&({\bf k}={\bf k}_1+{\bf k}_2+{\bf k}_3)\label{24}
\end{align}
\begin{align}
&R_5({\bf k}_1,\,{\bf k}_2,\,{\bf k}_3,\,{\bf k}_4)=NS_3({\bf k}_1,\,{\bf k}_2)S_2({\bf k}_3)\delta_{{\bf k}_3+{\bf k}_4}+NS_3({\bf k}_1,\,{\bf k}_3)S_2({\bf k}_2)\delta_{{\bf k}_2+{\bf k}_4}\notag\\
&+NS_3({\bf k}_1,\,{\bf k}_4)S_2({\bf k}_2)\delta_{{\bf k}_2+{\bf k}_3}+NS_3({\bf k}_2,\,{\bf k}_3)S_2({\bf k}_1)\delta_{{\bf k}_1+{\bf k}_4}+NS_3({\bf k}_2,\,{\bf k}_4)S_2({\bf k}_1)\delta_{{\bf k}_1+{\bf k}_3}\notag\\
&+NS_3({\bf k}_3,\,{\bf k}_4)S_2({\bf k}_1)\delta_{{\bf k}_1+{\bf k}_2}+NS_2({\bf k}_1)S_3({\bf k}_3,\,{\bf k}_4)\delta_{{\bf k}-{\bf k}_1}+NS_2({\bf k}_2)S_3({\bf k}_3,\,{\bf k}_4)\delta_{{\bf k}-{\bf k}_2}\notag\\
&+NS_2({\bf k}_3)S_3({\bf k}_2,\,{\bf k}_4)\delta_{{\bf k}-{\bf k}_3}+NS_2({\bf k}_4)S_3({\bf k}_2,\,{\bf k}_3)\delta_{{\bf k}-{\bf k}_3}\delta_{{\bf k}-{\bf k}_4};\quad({\bf k}={\bf k}_1+{\bf k}_2+{\bf k}_3+{\bf k}_4)\label{25}
\end{align}
The quadratic FDT \eqref{21} was established some time ago by evaluating its dynamical counterpart (see (31) below) in the static (dc) limit [7(a),\,(b)]. This was followed by the derivation of the cubic FDT \eqref{22} following a functional derivative approach [7(c)]. To the best of our knowledge, \eqref{23} with \eqref{25} is reported here for the first time.

Clearly, the combinations of $S_2S_2$ and $S_2S_3$ pair clusters comprising $R_4$ and $R_5$, respectively, remain invariant under permutation of their wave vector arguments. That is, the symmetry rules \eqref{4} and \eqref{5}, that apply to $S_4$ and $S_5$, apply to $R_4$ and $R_5$ as well. From FDTs \eqref{22} and \eqref{23}, it then follows that
\begin{align}
\hat{\chi}({\bf k}_1,\,{\bf k}_2,\,{\bf k}_3)=\hat{\chi}(-{\bf k},\,{\bf k}_1,\,{\bf k}_2)=\hat{\chi}({\bf k}_3,\,-{\bf k},\,{\bf k}_1)=\hat{\chi}({\bf k}_2,\,{\bf k}_3,\,-{\bf k});\qquad \quad ({\bf k}={\bf k}_1+{\bf k}_2+{\bf k}_3)\label{26}
\end{align}
\begin{align}
&\hat{\chi}({\bf k}_1,\,{\bf k}_2,\,{\bf k}_3,\,{\bf k}_4)=\hat{\chi}(-{\bf k},\,{\bf k}_1,\,{\bf k}_2,\,{\bf k}_3)=\hat{\chi}({\bf k}_4,\,-{\bf k},\,{\bf k}_1,\,{\bf k}_2)=\hat{\chi}({\bf k}_3,\,{\bf k}_4,\,-{\bf k},\,{\bf k}_1)\notag\\
&\phantom{{\hat{\chi}({\bf k}_1,\,{\bf k}_2,\,{\bf k}_3,\,{\bf k}_4)}}=\hat{\chi}({\bf k}_2,\,{\bf k}_3,\,{\bf k}_4,\,-{\bf k});\qquad  \qquad \qquad \qquad \qquad \qquad\qquad \,\,({\bf k}={\bf k}_1+{\bf k}_2+{\bf k}_3+{\bf k}_4)\label{27}
\end{align}
Looking further into the FDT hierarchy, we see that $R_6$ comrpises $S_2S_2S_2$, $S_2S_4$, $S_3S_3$ clusters; $R_7$ comprises $S_2S_5$, $S_3S_4$, $S_2S_2S_3$ clusters; and so on.
\\
\section{Hierarchy of Dynamical Fluctuation Dissipation Theorems}
We turn now to the formulation of the dynamical cubic ($p=3$) and quartic $(p=4)$ FDTs, each in a form that features one and only one $(p+1)$-point dynamical structure function
\begin{align}
&2\pi NS_{p+1}({\bf k}_1,\,{\bf k}_2,\,...,\,{\bf k}_p;\,\omega_1,\,\omega_2,\,...,\,\omega_p)\delta_{{\bf k}_1+{\bf k}_2+...+{\bf k}_p-{\bf k}}\delta(\omega_1+\omega_2+...+\omega_p-\omega)\notag\\
&=\langle n({\bf k}_1,\,\omega_1)n({\bf k}_2,\,\omega_2)...n({\bf k}_p,\,\omega_p)n(-{\bf k},\,-\omega)\rangle^{(0)},\quad ({\bf k}_1\not =0,\,{\bf k}_2\not =0,\,...,\,{\bf k}_p\not =0)\label{28}
\end{align}
expresed as a ${\bf k}\omega$-permutation ring combination of the $p$th-order density response functions defined through the constitutive relation
\begin{align}
\langle n({\bf k},\,\omega)\rangle^{(p)}=&\frac{1}{(2\pi V)^{p-1}}\sum_{{\bf k}_1}\sum_{{\bf k}_2}...\sum_{{\bf k}_p}\int_{-\infty}^\infty d\omega_1 \int_{-\infty}^\infty d\omega_2...\int_{-\infty}^\infty d\omega_p \hat{\chi}({\bf k}_1,\,{\bf k}_2,\,...,\,{\bf k}_p;\,\omega_1,\,\omega_2,\,...,\,\omega_p)\notag\\
&\times \hat{U}({\bf k}_1,\,\omega_1)\hat{U}({\bf k}_2,\,\omega_2)...\hat{U}({\bf k}_p,\,\omega_p)\delta_{{\bf k}_1+{\bf k}_2+...+{\bf k}_p-{\bf k}}\delta(\omega_1+\omega_2+...+\omega_p-\omega)\label{29}
\end{align}
Our derivation circumvents the need to proceed via the conventional Kubo approach [7(a),7(b), 8] which, at the levels of the cubic and quartic FDTs, would indeed be a daunting task. There are three structural guidelines that are key to our development of the dynamical hierarchy: i) the invariance of each FDT under permutations of its wave vector-frequency arguments; ii) the right-hand side (r.h.s.) structures of the static FDTs \eqref{20}-\eqref{23}, and iii) the left-hand-side (l.h.s.) structures of the dynamical linear and quadratic [7(a), 7(b)] FDTs:
\begin{align}
\Im i^0\bigg[ \frac{\hat{\chi}({\bf k}_1,\,\omega_1)}{\omega_1}-\frac{\hat{\chi}(-{\bf k}_1,\,-\omega_1)}{\omega_1}\bigg]=-\beta n_0 S_2({\bf k}_1,\,\omega_1)\label{30}
\end{align}
\begin{align}
&\Im i\bigg[
\frac{\hat{\chi}({\bf k}_1,\,{\bf k}_2;\,\omega_1,\,\omega_2)}{\omega_1\omega_2}
-\frac{\hat{\chi}(-{\bf k},\,{\bf k}_1;\,-\omega,\,\omega_1)}{\omega\omega_1}
-\frac{\hat{\chi}({\bf k}_2,\,-{\bf k};\,\omega,\,\omega_2)}{\omega_2\omega}
\bigg]
=-\frac{\beta^2 n_0}{2!2}S_3({\bf k}_1,\,{\bf k}_2;\,\omega_1,\,\omega_2)\notag\\
&\hspace{100mm}({\bf k}={\bf k}_1+{\bf k}_2,\quad \omega =\omega_1+\omega_2)\label{31}
\end{align}
The three guidelines above suggest the following structures for the cubic and quartic FDTs: 
\\
\begin{align}
&\Im i^2 \bigg[ 
\frac{\hat{\chi}({\bf k}_1,\,{\bf k}_2,\,{\bf k}_3;\,\omega_1,\,\omega_2,\,\omega_3)}{\omega_1\omega_2\omega_3}
-\frac{\hat{\chi}(-{\bf k},\,{\bf k}_1,\,{\bf k}_2;\,-\omega,\,\omega_1,\,\omega_2)}{\omega\omega_1\omega_2}
-\frac{\hat{\chi}({\bf k}_3,\,-{\bf k},\,{\bf k}_1;\,\omega_3,\,-\omega,\,\omega_1)}{\omega_3\omega\omega_1}\notag\\
&
-\frac{\hat{\chi}({\bf k}_2,\,{\bf k}_3,\,-{\bf k};\,\omega_2,\,\omega_3,\,-\omega)}{\omega_2\omega_3\omega}
\bigg]=-K_3 \bigg[ S_4({\bf k}_1,\,{\bf k}_2,\,{\bf k}_3;\,\omega_1,\,\omega_2,\,\omega_3)-R_4({\bf k}_1,\,{\bf k}_2,\,{\bf k}_3;\,\omega_1,\,\omega_2,\,\omega_3)\bigg]\notag\\
&\hspace{86mm} ({\bf k}={\bf k}_1+{\bf k}_2+{\bf k}_3;\quad \omega=\omega_1+\omega_2+\omega_3)\label{32}
\end{align}
\begin{align}
&\Im i^3 \bigg[ 
\frac{\hat{\chi}({\bf k}_1,\,{\bf k}_2,\,{\bf k}_3,\,{\bf k}_4;\,\omega_1,\,\omega_2,\,\omega_3,\,\omega_4)}{\omega_1\omega_2\omega_3\omega_4}
-\frac{\hat{\chi}(-{\bf k},\,{\bf k}_1,\,{\bf k}_2,\,{\bf k}_3;\,-\omega,\,\omega_1,\,\omega_2,\,\omega_3)}{\omega\omega_1\omega_2\omega_3}\notag\\
&\phantom{{\Im i^3\bigg[}}-\frac{\hat{\chi}({\bf k}_4,\,-{\bf k},\,{\bf k}_1,\,{\bf k}_2;\,\omega_4,\,-\omega,\,\omega_1,\,\omega_2)}{\omega_4\omega \omega_1\omega_2}
-\frac{\hat{\chi}({\bf k}_3,\,{\bf k}_4,\,-{\bf k},\,{\bf k}_1;\,\omega_3,\,\omega_4,\,-\omega,\,\omega_1)}{\omega_3\omega_4 \omega\omega_1}\notag\\
&\phantom{{\Im i^3\bigg[-\frac{\hat{\chi}({\bf k}_4,\,-{\bf k},\,{\bf k}_1,\,{\bf k}_2;\,\omega_4,\,-\omega,\,\omega_1,\,\omega_2)}{\omega_4\omega \omega_1\omega_2}}}
-\frac{\hat{\chi}({\bf k}_2,\,{\bf k}_3,\,{\bf k}_4,\,-{\bf k};\,\omega_2,\,\omega_3,\,\omega_4,\,-\omega)}{\omega_2\omega_3 \omega_4\omega}\bigg]\notag\\
&=-K_4\bigg[   
S_5({\bf k}_1,\,{\bf k}_2,\,{\bf k}_3,\,{\bf k}_4;\,\omega_1,\,\omega_2,\,\omega_3,\,\omega_4)-R_5({\bf k}_1,\,{\bf k}_2,\,{\bf k}_3,\,{\bf k}_4;\,\omega_1,\,\omega_2,\,\omega_3,\,\omega_4)
\bigg]\notag\\
&\hspace{70mm}({\bf k}={\bf k}_1+{\bf k}_2+{\bf k}_3+{\bf k}_4;\quad \omega=\omega_1+\omega_2+\omega_3+\omega_4)\label{33}
\end{align}
\\
\begin{align}
&R_4({\bf k}_1,\,{\bf k}_2,\,{\bf k}_3;\,\omega_1,\,\omega_2,\,\omega_3)=2\pi N\bigg[ S_2({\bf k}_1;\,\omega_1)S_2({\bf k}_2;\,\omega_2)\delta_{{\bf k}-{\bf k}_1}\delta(\omega-\omega_1)\notag\\
&+S_2({\bf k}_1;\,\omega_1)S_2({\bf k}_3;\,\omega_3)\delta_{{\bf k}-{\bf k}_3}\delta(\omega-\omega_3)+S_2({\bf k}_2;\,\omega_2)S_2({\bf k}_3;\,\omega_3)\delta_{{\bf k}-{\bf k}_2}\delta(\omega-\omega_2)\bigg]\label{34}
\end{align}
\\
\begin{align}
&R_5({\bf k}_1,\,{\bf k}_2,\,{\bf k}_3,\,{\bf k}_4;\,\omega_1,\,\omega_2,\,\omega_3,\,\omega_4)=2\pi N\bigg[ S_3({\bf k}_1,\,{\bf k}_2;\,\omega_1,\,\omega_2)S_2({\bf k}_3;\,\omega_3)\delta_{{\bf k}_3+{\bf k}_4}\delta(\omega_3+\omega_4)\notag\\
&+S_3({\bf k}_1,\,{\bf k}_3;\,\omega_1,\,\omega_3)S_2({\bf k}_2;\,\omega_2)\delta_{{\bf k}_2+{\bf k}_4}\delta(\omega_2+\omega_4)
+S_3({\bf k}_1,\,{\bf k}_4;\,\omega_1,\,\omega_4)S_2({\bf k}_2;\,\omega_2)\delta_{{\bf k}_2+{\bf k}_3}\delta(\omega_2+\omega_3)\notag\\
&+S_3({\bf k}_2,\,{\bf k}_3;\,\omega_2,\,\omega_3)S_2({\bf k}_1;\,\omega_1)\delta_{{\bf k}_1+{\bf k}_4}\delta(\omega_1+\omega_4)
+S_3({\bf k}_2,\,{\bf k}_4;\,\omega_2,\,\omega_4)S_2({\bf k}_1;\,\omega_1)\delta_{{\bf k}_1+{\bf k}_3}\delta(\omega_1+\omega_3)\notag\\
&+S_3({\bf k}_3,\,{\bf k}_4;\,\omega_3,\,\omega_4)S_2({\bf k}_1;\,\omega_1)\delta_{{\bf k}_1+{\bf k}_2}\delta(\omega_1+\omega_2)
+S_2({\bf k}_1;\omega_1)S_3({\bf k}_3,\,{\bf k}_4;\,\omega_3,\,\omega_4)\delta_{{\bf k}-{\bf k}_1}\delta(\omega-\omega_1)\notag\\
&+S_2({\bf k}_2;\,\omega_2)S_3({\bf k}_3,\,{\bf k}_4;\,\omega_3,\,\omega_4)\delta_{{\bf k}-{\bf k}_2}\delta(\omega-\omega_2)+S_2({\bf k}_3;\,\omega_3)S_3({\bf k}_2,\,{\bf k}_4;\,\omega_2,\,\omega_4)\delta_{{\bf k}-{\bf k}_3}\delta(\omega-\omega_3)\notag\\
&+S_2({\bf k}_4;\,\omega_4)S_3({\bf k}_2,\,{\bf k}_3;\,\omega_2,\,\omega_3)\delta_{{\bf k}-{\bf k}_4}\delta(\omega-\omega_4)\bigg]\label{35}
\end{align}
We next determine the constants $K_4$ and $K_5$. This is accomplished by integrating FDT equations \eqref{32} and \eqref{33} over those frequencies that are featured in the arguments of their r.h.s. structure functions $S_4$ and $S_5$; this action of course generates the $S_4({\bf k}_1,\,{\bf k}_2,\,{\bf k}_3)$ and $S_5({\bf k}_1,\,{\bf k}_2,\,{\bf k}_3,\,{\bf k}_4)$ ($t=0$) static structure functions featured in \eqref{22} and \eqref{23}. On the left-hand side, the reponse function integrals that ensue have removable singularities arising from the frequency factors in their denominators. Paralleling the procedure of [7(a)], such a combination of singular integrals involving causal response functions can be reformulated into a combination of Cauchy principle part integrals, amenable to Hilbert transform operations. Repeated applications of Kramers-Kronig formulas and Poincar\'e-Bertrand theorems, suitably generalized to handle multiple integrals, then bring \eqref{32} and \eqref{33} into forms which are identical to their companion static FDTs \eqref{22} and \eqref{23} for $K_3=\beta^3 n_0/(3!2^2)$ and $K_4=\beta^4 n_0/(4!2^3)$. We note that this is entirely consistent with the constants $K_1=\beta n_0/(1!2^0)$ and $K_2=\beta^2 n_0/(2!2)$ for the linear and quadratic FDTs \eqref{30} and \eqref{31}, respectively. Evidently, the constant for the $p$th member of the FDT hierarchy is given by $K_p=\beta^p n_0/(p!2^{p-1})$. The dynamical cubic and quartic FDTs \eqref{32} and \eqref{33} are the centerpieces of this paper.
\\
\section{Concluding Remarks}

Using a straightforward procedure that entirely circumvents the issue of the nested Poisson brackets encountered in the conventional Kubo formalism, we have derived tractable cubic $(p=3)$ and quartic $(p=4)$ FDTs, in which a single $(p+1)$-point dynamical structure function, entirely free of entangle Liouville space paths, is for the most part expressed as a linear combination of $p$th-order density response functions; these latter, in turn, can be readily calculated from model-dependent kinetic equations vis-\'a-vis constitutive relations linking the external and screened density response functions for the OCP. For $p\ge 3$, we note the emergence of the "remainder" terms $R_4$ and $R_5$, comprising of lower-order structure function pair clusters, which can be readily traded for response function clusters via the linear and quadratic FDTs \eqref{30} and \eqref{31}. Our study provides a clearer insight into the structure of the FDT hierarchy.

\end{document}